\documentclass[twocolumn,showpacs,preprintnumbers,amsmath,amssymb,nofootinbib,floatfix]{revtex4}
\usepackage[dvips]{graphicx}% Include figure files
\usepackage{dcolumn}% Align table columns on decimal point
\usepackage{bm}% bold math
\def\mapgeq{\mathbin{\lower.3ex\hbox{$\buildrel>\over{\smash{\scriptstyle\sim}\vphantom{_x}}$}}}
\def\mapleq{\mathbin{\lower.3ex\hbox{$\buildrel<\over{\smash{\scriptstyle\sim}\vphantom{_x}}$}}}
\def\mapgeqeq{\mathbi{\lower.3ex\hbox{$\buildrel>\over{\smash{\scriptstyle\approx}\vphantom{_2}}$}}}
\def\mapleqeq{\mathbin{\lower.3ex\hbox{$\buildrel<\over{\smash{\scriptstyle\approx}\vphantom{_2}}$}}}
\mathchardef\hanaO="724F
\def\Journal#1#2#3#4{{#1} {\bf #2} (#4) #3}
\def\MPL{Mod. Phys. Lett. A}

\def\NPB{Nucl. Phys. B}

\def\NPSUPPL{Nucl. Phys. Proc. Suppl.}
\def\PLB{{Phys. Lett.} B}

\def\PRL{Phys. Rev. Lett.}
\def\RMP{Rev. Mod. Phys.}
\def\PRD{Phys. Rev. D}

\def\PTP{Prog. Theor. Phys.}
\def\JHEP{JHEP}
\def\JCAP{JCAP}
\def\NPBSUPPL{Nucl. Phys. B. Proc. Suppl.}
\def\EPJ{Euro. Phys. J. C}

\def\JETPUSSR{Sov. Phys. JETP}

\def\ZETP{Zh. Eksp. Teor. Fiz.}
\def\PismaZETP{Pis'ma Zh. Eksp. Teor. Fiz.}

\def\JPG{J. Phys. G}

\def\APJ{Astrophys. J.}
\def\APJS{Astrophys. J. Suppl}

\def\NJP{New J. Phys.}

\def\Erratum{Erratum-ibid}
\def\SCIENCE{Science}

\begin{document}

%\preprint{TOKAI-HEP/TH-0603}

 \title{Generalized Scaling Ansatz and Minimal Seesaw Mechanism}
% Force line breaks with \\

% \author{Teruyuki Kitabayashi}
% \email{teruyuki@keyaki.cc.u-tokai.ac.jp}
% \affiliation[Also at ]{Physics Department, XYZ University.}%Lines break automatically or can be forced with \\
% \affiliation{\vspace{3mm}%
% \sl Department of Science and Technology,\\
% Graduate School of Science and Technology, Tokai University\\
% 1117 Kitakaname, Hiratsuka, Kanagawa 259-1292, Japan\\
% }
%\author{Ryo Takase}
%\email{9asnm009@mail.tokai-u.jp}
\author{Masaki Yasu\`{e}}
\email{yasue@keyaki.cc.u-tokai.ac.jp}
\affiliation{\vspace{3mm}%
\sl Department of Physics, Tokai University,\\
4-1-1 Kitakaname, Hiratsuka, Kanagawa 259-1292, Japan\\
}

%\date{October, 2012}% It is always \today, today,
             %  but any date may be explicitly specified

%%-------------------------------------------------
%% Abstract
%%-------------------------------------------------
\begin{abstract}
Generalized scaling in flavor neutrino masses $M_{ij}$ ($i,j$=$e,\mu,\tau$) expressed in terms of $\theta_{SC}$ and the atmospheric neutrino mixing angle $\theta_{23}$ is defined by $M_{i\tau }/M_{i\mu }$ = $- \kappa _it_{23}$ ($i$=$e,\mu,\tau$) with $\kappa _e$=1, $\kappa _\mu$=$B/A$ and $\kappa _\tau$=$1/B$, where $t_{23}=\tan\theta_{23}$, $A$=${\cos ^2}{\theta _{SC}}+{\sin ^2}{\theta _{SC}}t_{23}^4$ and $B$=${\cos ^2}{\theta _{SC}}-{\sin ^2}{\theta _{SC}}t_{23}^2$.  The generalized scaling ansatz predicts the vanishing reactor neutrino mixing angle $\theta_{13}=0$.  It is shown that the minimal seesaw mechanism naturally implements our scaling ansatz. There are textures satisfying the generalized scaling ansatz that yield vanishing baryon asymmetry of the Universe (BAU).  Focusing on these textures, we discuss effects of $\theta_{13}\neq 0$ to evaluate a CP-violating Dirac phase $\delta$ and BAU and find that BAU is approximately controlled by the factor $\sin^2\theta_{13}\sin(2\delta -\phi)$, where $\phi$ stands for the CP-violating Majorana phase whose magnitude turns out to be at most 0.1. 
\end{abstract}

% PACS, the Physics and Astronomy
                             % Classification Scheme.
%\keywords{Keywords: neutrino mass, radiative mechanism, lepton triplet}%Use showkeys class option if keyword
                              %display desired
\maketitle
%%-------------------------------------------------
%% Main body
%%-------------------------------------------------
\section{\label{sec:intro} Introduction}
The recent observation of the nonvanishing reactor mixing angle $\theta_{13}$ \cite{theta13} opens a new window to clarify properties of CP violation in neutrino physics.  CP violation occurs in neutrino oscillations \cite{CPViolation} and in leptogenesis \cite{Leptogenesis} based on the seesaw mechanism \cite{SeeSaw}. In the seesaw mechanism, neutrinos are almost Majorana particles generated by heavy Majorana neutrinos as heavy as ${\mathcal O}(10^{10})$ GeV and turn out to be extremely light so that they are compatible with experimental observations \cite{atmospheric,accelerator, oldsolar,solar,reactor}.  Effects of CP violation in the lighter Majorana neutrinos are characterized by phases of the Pontecorvo-Maki-Nakagawa-Sakata (PMNS) unitary matrix $U_{PMNS}$ \cite{PMNS}, which converts massive neutrinos $\nu_{1,2,3}$ into flavor neutrinos $\nu_{e,\mu,\tau}$. Three phases, one CP-violating Dirac phase $\delta$ and two CP-violating Majorana phases $\phi_{2,3}$ are involved in $U_{PMNS}$ \cite{CPViolation}. On the other hand,  CP violation in leptogenesis is characterized by phases related to heavy Majorana neutrinos.  These two types of CP violation are, in principle, independent of each other.  However, they can be correlated if there are some constraints that reduce the number of degrees of freedom, which result in relating two different types of CP-phases.  It is known that the minimal seesaw mechanism utilizing two heavy neutrinos \cite{MinimalSeesaw} involves three physical CP-violating phases in leptogenesis, which are equivalent to $\delta$ and $\phi_{2,3}$; therefore, CP violation in leptogenesis can be controlled by $\delta$ and $\phi_{2,3}$.

The observed $\sin^2\theta_{13}$ is found to be $\sin^2\theta_{13}\approx 0.025$ \cite{NuData, NuData2} close to $\sin^2\theta_{13}= 0$, which suggests a theoretical principle that $\sin^2\theta_{13}$ vanishes as the first approximation and that a certain perturbation induces nonvanishing $\sin^2\theta_{13}$.  There have been various theoretical  ideas that give $\sin^2\theta_{13}=0$ \cite{Bimaximal, Tribimaximal, transposedTribimaximal, Bipair, goldenRatio, hexagonal, mu-tau, theta_13=0, Scaling, GeneralizedScaling}.  Among others, the generalized scaling ansatz in flavor neutrino masses, which is an extended version of the scaling ansatz \cite{Scaling}, is proposed to discuss a new aspect of neutrinos \cite{GeneralizedScaling}.  The generalized scaling is described by two angles, $\theta_{SC}$ and the atmospheric neutrino mixing angle $\theta_{23}$, which provide the following scaling rule among Majorana flavor neutrino masses $M_{ij}$ ($i,j=e,\mu,\tau$):
%%%%%%%%%%%%%%%%%%%%
\begin{eqnarray}
\frac{{{M_{i\tau }}}}{{{M_{i\mu }}}} =  - {\kappa _i}{t_{23}}~\left(i=e,\mu,\tau\right),
\label{Eq:ScalingRule}
\end{eqnarray}
%%%%%%%%%%%%%%%%%%%%
where $t_{23}=\tan\theta_{23}$, $\left(\kappa_e, \kappa_\mu, \kappa_\tau\right)$=$\left( 1,B/A,1/B\right)$ and 
%%%%%%%%%%%%%%%%%%%%
\begin{eqnarray}
A &=& {\cos ^2}{\theta _{SC}} + {\sin ^2}{\theta _{SC}}t_{23}^4,
\nonumber\\
B &=& {\cos ^2}{\theta _{SC}} - {\sin ^2}{\theta _{SC}}t_{23}^2.
\label{Eq:A-B}
\end{eqnarray}
%%%%%%%%%%%%%%%%%%%%
It can be proved that Eq.(\ref{Eq:ScalingRule}) indices $\theta_{13}=0$. The condition to obtain $\theta_{13}=0$ consists of the following two relations \cite{theta_13=0}:
%%%%%%%%%%%%%%%%%%%%
\begin{eqnarray}
{M_{e\tau }} &=& - {t_{23}}{M_{e\mu }},
\label{Eq:Constraints-Metau}\\
{M_{\tau \tau }} &=& {M_{\mu \mu }} + \frac{{1 - t_{23}^2}}{{{t_{23}}}}{M_{\mu \tau }},
\label{Eq:Constraints-Mtautau}
\end{eqnarray}
%%%%%%%%%%%%%%%%%%%%
where Eq.(\ref{Eq:ScalingRule}) turns out to satisfy these relations and the generalized scaling maintains $\theta_{13}=0$.  The angle $\theta_{SC}$ itself is defined from ${{{M_{\mu\tau }}}}/{{{M_{\mu\mu }}}} =  - {\kappa _\mu}{t_{23}}$ to be \cite{GeneralizedScaling}:
%%%%%%%%%%%%%%%%%%%%
\begin{eqnarray}
{{\sin }^2}{\theta _{SC}} = \frac{{c_{23}^2\left( {{M_{\mu \tau }} + {t_{23}}{M_{\mu \mu }}} \right)}}{{\left( {1 - t_{23}^2} \right){M_{\mu \tau }} + {t_{23}}{M_{\mu \mu }}}},
\label{Eq:Sin2SC}
\end{eqnarray}
%%%%%%%%%%%%%%%%%%%%
where $c_{23}=\cos\theta_{23}$.

In this article, we would like to demonstrate that the generalized scaling rule is naturally realized in the minimal seesaw mechanism and to discuss the creation of the baryon asymmetry of the Universe (BAU) via leptogenesis \cite{ScalingMinimalSeesaw}.  In Sec.\ref{sec:seesaw}, some of seesaw textures that satisfy the generalized scaling ansatz are found to yield the vanishing BAU. In these textures, it is expected that breaking effects of the generalized scaling ansatz initiate creating BAU and simultaneously inducing Dirac CP violation as a result of $\theta_{13}\neq 0$.  In Sec.\ref{sec:leptogenesis}, we describe leptogenesis based on the minimal seesaw mechanism and show theoretical arguments to make predictions on possible correlations between BAU and CP-violating phases.  In Sec.\ref{sec:numerical}, a numerical analysis is performed to estimate sizes of BAU and of Dirac and Majorana CP-violations, which will be compared with our theoretical predictions. The final section, Sec.\ref{sec:summary}, is devoted to a summary.

\section{\label{sec:seesaw} Seesaw Textures}
The minimal seesaw mechanism introduces two heavy Majorana neutrinos $N_{1,2}$ into the standard model. We understand that a $2 \times 2$ heavy neutrino mass matrix $M_R$ and a charged lepton mass matrix are transformed into diagonal and real ones.  After the heavy neutrinos are decoupled, the minimal seesaw mechanism generates a symmetric $3 \times 3$ light neutrino mass matrix $M_\nu$ given by $M_\nu = -m_D M_R^{-1} m_D^T$, where $m_D$ is a $3 \times 2$ Dirac neutrino mass matrix. We parametrize $M_R$ by 
%%%%%%%%%%%%%%%%%%%%
\begin{eqnarray}
M_R = \left(
  \begin{array}{cc}
  M_1 & 0   \\ 
  0       & M_2
  \end{array}
\right)\quad (M_2>M_1),
\end{eqnarray}
%%%%%%%%%%%%%%%%%%%%
and $m_D$ by
%%%%%%%%%%%%%%%%%%%%
\begin{eqnarray}
m_D = 
\left(
  \begin{array}{cc}
    \sqrt{M_1}a_1  & \sqrt{M_2}b_1   \\
    \sqrt{M_1}a_2  & \sqrt{M_2}b_2   \\
    \sqrt{M_1}a_3  & \sqrt{M_2}b_3   \\
  \end{array}
\right),
\end{eqnarray}
%%%%%%%%%%%%%%%%%%%%
which results in
%%%%%%%%%%%%%%%%%%%%
\begin{eqnarray}
M_\nu &=& \left( \begin{array}{*{20}{c}}
M_{ee}&M_{e\mu }&M_{e\tau }\\
M_{e\mu }&M_{\mu \mu }&M_{\mu \tau }\\
M_{e\tau }&M_{\mu \tau }&M_{\tau \tau }
\end{array} \right)
\nonumber\\
&=&
-\left(
  \begin{array}{ccc}
    a_1^2 + b_1^2   & a_1a_2 + b_1b_2   &  a_1a_3 + b_1b_3  \\
    a_1a_2 + b_1b_2 & a_2^2 + b_2^2     &  a_2a_3 + b_2b_3  \\
    a_1a_3 + b_1b_3 & a_2a_3 + b_2b_3   &  a_3^2 + b_3^2   \\
  \end{array}
\right),
\nonumber\\
\label{Eq:Mnu}
\end{eqnarray}
%%%%%%%%%%%%%%%%%%%%
where the minus sign in front of the mass matrix is discarded for the later discussions.  One of the masses of $\nu_{1,2,3}$ is required to vanish owing to $\det\left(M_\nu\right)=0$. 

To obtain the seesaw version of the generalized scaling ansatz, we describe the basic conditions on $M_\nu$ Eqs.(\ref{Eq:Constraints-Metau}) and (\ref{Eq:Constraints-Mtautau}) in terms of seesaw mass parameters $a_{1,2,3}$ and $b_{1,2,3}$ and search their solutions in much the same way Eq.(\ref{Eq:ScalingRule}) is derived.  These conditions are readily converted into
%%%%%%%%%%%%%%%%%%%%
%\begin{widetext}
\begin{eqnarray}
&& 
\left( {{a_3} + {t_{23}}{a_2}} \right){a_1} + \left( {{b_3} + {t_{23}}{b_2}} \right){b_1} = 0,
\label{Eq:Constraints-mD_1}\\
&& 
\left( {{t_{23}}{a_3} - {a_2}} \right)\left( {{a_3}  + {t_{23}}{a_2}} \right) 
\nonumber\\
&& 
\qquad
+ \left( {{t_{23}}{b_3} - {b_2}} \right)\left( {{b_3} + {t_{23}}{b_2}} \right) = 0.
\label{Eq:Constraints-mD_2}
\end{eqnarray}
%\end{widetext}
%%%%%%%%%%%%%%%%%%%%
The minimal seesaw mechanism that keeps $\theta_{13}$ vanished should satisfy Eqs.(\ref{Eq:Constraints-mD_1}) and (\ref{Eq:Constraints-mD_2}).  The simpler solutions to Eqs.(\ref{Eq:Constraints-mD_1}) and (\ref{Eq:Constraints-mD_2}) can be either
%%%%%%%%%%%%%%%%%%%%
\begin{enumerate}
\item ${a_3} + {t_{23}}{a_2} = {b_3} + {t_{23}}{b_2} = 0$ 
leading to
%%%%%%%%%%%%%%%%%%%%
%\begin{widetext}
\begin{eqnarray}
{m_D} = \left( {\begin{array}{*{20}{c}}
{\sqrt {{M_1}} {a_1}}&{\sqrt {{M_2}} {b_1}}\\
{\sqrt {{M_1}} {a_2}}&{\sqrt {{M_2}} {b_2}}\\
{\sqrt {{M_1}} \left( { - {t_{23}}{a_2}} \right)}&{\sqrt {{M_2}} \left( { - {t_{23}}{b_2}} \right)}
\end{array}} \right),
\label{Eq:Solution_mD_1}
\end{eqnarray}
%\end{widetext}
%%%%%%%%%%%%%%%%%%%%
or
\item $a_1 = 0$, $t_{23}a_3 - a_2 = 0$ and $b_3 + t_{23}b_2 = 0$,
leading to
%%%%%%%%%%%%%%%%%%%%
%\begin{widetext}
\begin{eqnarray}
{m_D} = \left( {\begin{array}{*{20}{c}}
{0}&{\sqrt {{M_2}} {b_1}}\\
{\sqrt {M_1}} {a_2}&{\sqrt {{M_2}} {b_2}}\\
{\sqrt {M_1}} a_2/t_{23}&{\sqrt {{M_2}} \left( { - {t_{23}}{b_2}} \right)}
\end{array}} \right),
\label{Eq:Solution_mD_2}
\end{eqnarray}
%\end{widetext}
%%%%%%%%%%%%%%%%%%%%
and a solution with $b_1=0$ is 
%%%%%%%%%%%%%%%%%%%%
%\begin{widetext}
\begin{eqnarray}
{m_D} = \left( {\begin{array}{*{20}{c}}
{\sqrt {{M_1}} {a_1}}&{0}\\
{\sqrt {M_1}} {a_2}&{\sqrt {{M_2}} {b_2}}\\
{\sqrt {M_1} \left( { - {t_{23}}{a_2}} \right)}&{\sqrt {M_2}} b_2/t_{23}
\end{array}} \right).
\label{Eq:Solution_mD_3}
\end{eqnarray}
%\end{widetext}
%%%%%%%%%%%%%%%%%%%%
\end{enumerate}
%%%%%%%%%%%%%%%%%%%%
Although there are other solutions,\footnote{A solution can be supplied by ${b_1} =  - {t_{23}}{a_1}$, ${b_2} = t_{23}^2/a_3$ and ${b_3} = {a_2}$ with $t^2_{23}=1$, which describe a $\mu$-$\tau$ symmetric seesaw model \cite{MuTauSeasaw}.} the above solutions suffice to show consistent results with the generalized scaling ansatz. 

For Eq.(\ref{Eq:Solution_mD_1}) in the case 1, we find that
%%%%%%%%%%%%%%%%%%%%
%\begin{widetext}
\begin{eqnarray}
{M_{e\tau }} &=&  - {t_{23}}\left( {{a_1}{a_2} + {b_1}{b_2}} \right)=-t_{23} M_{e\mu},
\label{Eq:Solution_etau_ab1}
\end{eqnarray}
%\end{widetext}
%%%%%%%%%%%%%%%%%%%%
and
%%%%%%%%%%%%%%%%%%%%
%\begin{widetext}
\begin{eqnarray}
{M_{\mu \mu }} &=& a_2^2 + b_2^2,
\quad
{M_{\mu \tau }} =  - t_{23}\left(a_2^2 + b_2^2\right),
\nonumber\\
{M_{\tau \tau }} &=& t^2_{23}\left(a_2^2 + b_2^2\right),
\label{Eq:Solution_ab1}
\end{eqnarray}
%\end{widetext}
%%%%%%%%%%%%%%%%%%%%
from which Eq.(\ref{Eq:Sin2SC}) leads to
%%%%%%%%%%%%%%%%%%%%
\begin{eqnarray}
\sin^2\theta_{SC} = 0,
\label{Eq:Solution_ab1_SC}
\end{eqnarray}
%%%%%%%%%%%%%%%%%%%%
corresponding to the inverted mass hierarchy with $m_3=0$ \cite{Scaling}.  On the other hand, for Eq.(\ref{Eq:Solution_mD_2}) in the case 2, we find that
%%%%%%%%%%%%%%%%%%%%
%\begin{widetext}
\begin{eqnarray}
{M_{e\tau }} &=&  - {t_{23}}{b_1}{b_2}=-t_{23} M_{e\mu},
\label{Eq:Solution_etau_ab2}
\end{eqnarray}
%\end{widetext}
%%%%%%%%%%%%%%%%%%%%
and
%%%%%%%%%%%%%%%%%%%%
%\begin{widetext}
\begin{eqnarray}
{M_{\mu \mu }} &=& a_2^2 + b_2^2,
\quad
{M_{\mu \tau }} = \frac{1}{{{t_{23}}}}a_2^2 - {t_{23}}b_2^2,
\nonumber\\
{M_{\tau \tau }} &=& \frac{1}{{{t^2_{23}}}}a_2^2 + {t^2_{23}}b_2^2,
\label{Eq:Solution_ab2}
\end{eqnarray}
%\end{widetext}
%%%%%%%%%%%%%%%%%%%%
from which Eq.(\ref{Eq:Sin2SC}) leads to
%%%%%%%%%%%%%%%%%%%%
%\begin{widetext}
\begin{eqnarray}
{\sin ^2}{\theta _{SC}} &=& \frac{\left(a_2/t_{23}\right)^2}{\left(a_2/t_{23}\right)^2 + \left(t_{23}b_2\right)^2}.
\label{Eq:Solution_ab2_SC}
\end{eqnarray}
%\end{widetext}
%%%%%%%%%%%%%%%%%%%%
Similarly, we obtain that
%%%%%%%%%%%%%%%%%%%%
%\begin{widetext}
\begin{eqnarray}
{\sin ^2}{\theta _{SC}} &=& \frac{\left(b_2/t_{23}\right)^2}{\left(t_{23}a_2\right)^2+\left(b_2/t_{23}\right)^2},
\label{Eq:Solution_ab3_SC}
\end{eqnarray}
%\end{widetext}
%%%%%%%%%%%%%%%%%%%%
for Eq.(\ref{Eq:Solution_mD_3}).  These definitions of $\sin^2\theta_{SC}$ depending upon the types of seesaw textures provide the seesaw version of Eq.(\ref{Eq:Sin2SC}).

Since the case with $m_3=0$ is described by Eq.(\ref{Eq:Solution_mD_1}), Eqs.(\ref{Eq:Solution_mD_2}) and (\ref{Eq:Solution_mD_3}) should describe the normal mass hierarchy with $m_1$ = 0.  In other words, the inverted mass hierarchy with $m_3=0$ realized at $\theta_{13}\neq 0$ does not approach the ideal textures Eqs.(\ref{Eq:Solution_mD_2}) and Eq.(\ref{Eq:Solution_mD_3}) at $\theta_{13}=0$.  This is because, for $\theta_{13}\neq 0$, we obtain that, for an arbitrary parameter $x$,
%%%%%%%%%%%%%%%%%%%%
%\begin{widetext}
\begin{eqnarray}
&&
M_{\mu \mu } + 2x{M_{\mu \tau }} + {x^2}{M_{\tau \tau }} 
= {\left( {{a_2} + x{a_3}} \right)^2} + {\left( {{b_2} + x{b_3}} \right)^2},
\nonumber\\
&&
\label{Eq:Inverted_ab_Case1}
\end{eqnarray}
%\end{widetext}
%%%%%%%%%%%%%%%%%%%%
as well as
%%%%%%%%%%%%%%%%%%%%
%\begin{widetext}
\begin{eqnarray}
&&
M_{\mu \mu } + 2x{M_{\mu \tau }} + {x^2}{M_{\tau \tau }} 
\nonumber\\
&&
=\left( { - {c_{23}}{s_{12}} - {s_{23}}{c_{12}}{{\tilde s}_{13}} + x\left( {{s_{23}}{s_{12}} - {c_{23}}{c_{12}}\tilde s_{13}^\ast } \right)} \right)^2{\tilde m_1} 
\nonumber\\
&&
\quad
+ {\left( {{c_{23}}{c_{12}} - {s_{23}}{s_{12}}{{\tilde s}_{13}} - x\left( {{s_{23}}{c_{12}} + {c_{23}}{s_{12}}\tilde s_{13}^\ast } \right)} \right)^2}{\tilde m_2},
\nonumber\\
&&
\label{Eq:Inverted_Mij_Case1}
\end{eqnarray}
%\end{widetext}
%%%%%%%%%%%%%%%%%%%%
where $s_{ij} = \sin\theta_{ij}$ and $c_{ij} = \cos\theta_{ij}$ ($i,j$=1,2,3) for $\theta_{12}$ being the solar neutrino mixing angle, ${\tilde s_{13}} = s_{13}e^{i\delta}$ and ${\tilde m_a} = m_ae^{-i\varphi_a}$ ($a=1,2,3$) for $\varphi_a$ being Majorana phases.\footnote{The CP-violating Majorana phase $\phi$ is defined by $\phi=\varphi_3-\varphi_2$ for $m_1=0$ and $\phi=\varphi_2-\varphi_1$ for $m_3=0$.}  At $x = 1/t_{23}$, Eq.(\ref{Eq:Inverted_Mij_Case1}) vanishes if $\theta_{13}=0$; therefore, Eq.(\ref{Eq:Inverted_ab_Case1}) vanishes as well. On the other hand, at $x = 1/t_{23}$, it is Eq.(\ref{Eq:Solution_mD_1}) that gives the vanishing of Eq.(\ref{Eq:Inverted_ab_Case1}).  At $\theta_{13}=0$, Eq.(\ref{Eq:Solution_mD_1}) is, thus, derived.  When $m_3\neq 0$, such as in a seesaw mechanism with a 3$\times$3 $m_D$, is taken, Eqs.(\ref{Eq:Solution_mD_2}) and (\ref{Eq:Solution_mD_3}) can describe the inverted mass hierarchy.

\section{\label{sec:leptogenesis} CP Violation and Leptogenesis}
Leptogenesis creates BAU whose estimate contains the factor $(m_D^\dag {m_D})_{12}$ \cite{Leptogenesis} which turns out to be $a_1^\ast b_1 + a_2^\ast b_2 + a_3^\ast b_3$.  It is found that BAU vanishes for the seesaw textures, Eqs.(\ref{Eq:Solution_mD_2}) and (\ref{Eq:Solution_mD_3}), which yield $a_1^\ast b_1 + a_2^\ast b_2 + a_3^\ast b_3=0$ \cite{ZeroPhotonBaryon}.  If these textures of $m_D$ are adopted, CP violation of leptogenesis and of the Dirac type for flavor neutrinos becomes active only if sources of $\theta_{13}\neq 0$ are present \cite{DiracPhaseLeptogenesis}.  For the rest of discussions, we focus our attention on these seesaw textures to discuss how the creation of BAU relates to CP violation for flavor neutrinos.  We restrict ourselves to discussions based on Eq.(\ref{Eq:Solution_mD_2}), from which results from Eq.(\ref{Eq:Solution_mD_3}) can be obtained by the interchange of $a_{1,2,3}\leftrightarrow b_{1,2,3}$ unless otherwise specified.  

To obtain $\theta_{13}\neq 0$ and the nonvanishing BAU, we include breaking terms of the generalized scaling ansatz, which are denoted  by $\delta a_3$ and $\delta b_3$ to give
%%%%%%%%%%%%%%%%%%%%
%\begin{widetext}
\begin{eqnarray}
{a_3} = \frac{{{a_2}}}{{{t_{23}}}} + \delta {a_3},
\quad
{b_3} =  - {t_{23}}{b_2} + \delta {b_3}.
\label{Eq:delta a3-b3}
\end{eqnarray}
%\end{widetext}
%%%%%%%%%%%%%%%%%%%%
The angle $\theta_{SC}$ is still defined by Eq.(\ref{Eq:Sin2SC}) and the $i=\tau$ part of the generalized scaling rule Eq.(\ref{Eq:ScalingRule}) gets broken according to
%%%%%%%%%%%%%%%%%%%%
%\begin{widetext}
\begin{eqnarray}
&&
M_{\tau\tau }+ \kappa _\tau{t_{23}}M_{\mu\tau }  = \left( {{a_3} + {t_{23}}{a_2}} \right)\delta {a_3} + \left( {{b_3} - \frac{{{b_2}}}{{{t_{23}}}}} \right)\delta {b_3}.
\nonumber\\
&&
\label{Eq:Scaling_ab}
\end{eqnarray}
%\end{widetext}
%%%%%%%%%%%%%%%%%%%%
The nonvanishing BAU is generated owing to $a_1^\ast b_1 + a_2^\ast b_2 + a_3^\ast b_3\neq 0$, which is calculated to be:
%%%%%%%%%%%%%%%%%%%%
%\begin{widetext}
\begin{eqnarray}
a_1^\ast{b_1} + a_2^\ast{b_2} + a_3^\ast{b_3} &=& \frac{1}{{{t_{23}}}}a_2^\ast\delta {b_3} - {t_{23}}\delta a_3^\ast{b_2} 
\nonumber\\
&&
+\delta a_3^\ast\delta b_3.
\label{Eq:PhtonToBaryonBreaking}
\end{eqnarray}
%\end{widetext}
%%%%%%%%%%%%%%%%%%%%
On the other hand, the CP-violating Dirac phase $\delta$, which is contained in a specific version of $U_{PMNS}$ defined by the Particle Data Group \cite{PDG}, is estimated to be  \cite{DiracCPFormula}
%%%%%%%%%%%%%%%%%%%%
%\begin{widetext}
\begin{eqnarray}
\delta  &=& \arg\left[\left( {\frac{1}{{{t_{23}}}}M_{\mu \tau }^\ast + M_{\mu \mu }^\ast + \delta M_{\tau \tau }^\ast} \right)\delta {M_{e\tau }} + {M_{ee}}\delta M_{e\tau }^\ast\right. 
\nonumber\\
&&
\left.
-{t_{23}}{M_{e\mu }}\delta {M_{\tau \tau }}\right],
\label{Eq:DeltaCP}
\end{eqnarray}
%\end{widetext}
%%%%%%%%%%%%%%%%%%%%
for the normal mass hierarchy, where $\delta M_{e\tau}$ and $\delta M_{\tau\tau}$ calculated from
%%%%%%%%%%%%%%%%%%%%
%\begin{widetext}
\begin{eqnarray}
\delta {M_{e\tau }} &=& {M_{e\tau }} + {t_{23}}{M_{e\mu }},
\nonumber\\
\delta {M_{\tau \tau }} &=& {M_{\tau \tau }} - \left( {{M_{\mu \mu }} + \frac{{1 - t_{23}^2}}{{{t_{23}}}}{M_{\mu \tau }}} \right),
\label{Eq:DeltaMetauMtautau}
\end{eqnarray}
%\end{widetext}
%%%%%%%%%%%%%%%%%%%%
turn out to be
%%%%%%%%%%%%%%%%%%%%
%\begin{widetext}
\begin{eqnarray}
\delta {M_{e\tau }} &=& {b_1}\delta {b_3},
\nonumber\\
\delta {M_{\tau \tau }} &=& \frac{{1 + t_{23}^2}}{{{t_{23}}}}\left({a_2}\delta {a_3} -{b_2}\delta {b_3} \right)
+ {\left( {\delta {a_3}} \right)^2} + {\left( {\delta {b_3}} \right)^2}.
\nonumber\\
&&
\label{Eq:DeltaMetauMtautauByab}
\end{eqnarray}
%\end{widetext}
At the same time, $\theta _{13}$ is calculated by the following formula \cite{FormulaTheta13}:
%%%%%%%%%%%%%%%%%%%%
%\begin{widetext}
\begin{eqnarray}
&&
\tan 2{\theta _{13}} 
\nonumber\\
&&
= \frac{{2{c_{23}}\delta {M_{e\tau }}}}{{\left( {s_{23}^2{M_{\mu \mu }} + c_{23}^2{M_{\tau \tau }} + 2{s_{23}}{c_{23}}{M_{\mu \tau }}} \right){e^{i\delta }} - {M_{ee}}{e^{ - i\delta }}}}.
\nonumber\\
\label{Eq:Theta13Exact}
\end{eqnarray}
%\end{widetext}
%%%%%%%%%%%%%%%%%%%%

We can further approximate Eqs.(\ref{Eq:PhtonToBaryonBreaking}),  (\ref{Eq:DeltaCP}) and (\ref{Eq:Theta13Exact}) to see that the breaking $\delta b_3$ is a main source to start creating BAU and inducing the nonvanishing $\delta$ and $\theta_{13}$. The normal mass hierarchy demands that $\left|M_{\mu\mu,\mu\tau,\tau\tau}\right|\gg\left|M_{ee,e\mu,e\tau}\right|$, which are equivalent to $\left|a^2_{2,3}\right|\gg\left|b^2_{1,2,3}\right|$.  For the region where second order terms with respect to $\delta a_3$ and $\delta b_3$ are safely neglected, we obtain that
%%%%%%%%%%%%%%%%%%%%
%\begin{widetext}
\begin{eqnarray}
&& 
a_1^\ast {b_1} + a_2^\ast {b_2} + a_3^\ast {b_3} \approx \frac{1}{{{t_{23}}}}a_2^\ast \delta {b_3},
\label{Eq:ApproximatedPhtonToBaryonBreaking}\\
&&
\delta  \approx \arg \left( a_2^{\ast 2}b_1\delta b_3 \right),
\label{Eq:ApproximatedDeltaCP}
\end{eqnarray}
%\end{widetext}
%%%%%%%%%%%%%%%%%%%%
as well as
%%%%%%%%%%%%%%%%%%%%
%\begin{widetext}
\begin{eqnarray}
\tan 2{\theta _{13}} \approx 2c_{23}s_{23}^2\left| {\frac{{{b_1}\delta {b_3}}}{{a_2^2}}} \right|,
\label{Eq:Theta13}
\end{eqnarray}
%\end{widetext}
%%%%%%%%%%%%%%%%%%%%
for $\sin ^2\theta _{SC} \approx 1$.  Therefore, we understand that the main source of CP-violations and $\theta_{13}\neq 0$ is the single breaking term $\delta b_3$.  It should be noted that, for the $b_1=0$ texture, 
%%%%%%%%%%%%%%%%%%%%
%\begin{widetext}
\begin{eqnarray}
&& 
a_1^\ast {b_1} + a_2^\ast {b_2} + a_3^\ast {b_3} \approx \frac{1}{{{t_{23}}}}b_2 \delta a_3^\ast,
\label{Eq:ApproximatedPhtonToBaryonBreaking_b1=0}
\end{eqnarray}
%\end{widetext}
%%%%%%%%%%%%%%%%%%%%
is derived in place of Eq.(\ref{Eq:ApproximatedPhtonToBaryonBreaking}).

We describe various seesaw parameters to estimate BAU from leptogenesis based on the minimal seesaw mechanism.  The recipes to calculate BAU are given as follows \cite{FormulaLeptogenesis,EstimateLeptogenesis}:
%%%%%%%%%%%%%%%%%%%%
\begin{enumerate}
%%%%%%%%%%%%%%%%%%%%
\item The heavy neutrinos are taken to satisfy the hierarchical mass pattern of $M_1\ll M_2$, where the CP-asymmetry from $N_2$ is washed out;
\item The CP-asymmetry from the decay of $N_1$ is given by the flavor-dependent ${\varepsilon ^\alpha }$ ($\alpha = e,\mu,\tau$):
%%%%%%%%%%%%%%%%%%%%
%\begin{widetext}
\begin{eqnarray}
&&
{\varepsilon ^\alpha } = \frac{1}{{8\pi {v^2}}}\frac{{{\rm Im} \left[ {{\left( {m_D^\dag } \right)}_{1\alpha }}{{\left( {{m_D}} \right)}_{\alpha 2}}\left( {m_D^\dag {m_D}} \right)_{12} \right]}}{{{{\left( {m_D^\dag {m_D}} \right)}_{11}}}}
\nonumber\\
&&
\quad
\cdot f\left( {\frac{{{M_2}}}{{{M_1}}}} \right),
\label{Eq:Epsilon}
\end{eqnarray}
%\end{widetext}
%%%%%%%%%%%%%%%%%%%%
where $v \simeq 174$ GeV and
%%%%%%%%%%%%%%%%%%%%
%\begin{widetext}
\begin{eqnarray}
&&
f\left( x \right) 
\nonumber\\
&&
= x\left[ {1 - \left( {1 + {x^2}} \right)\ln \left( {\frac{{1 + {x^2}}}{{{x^2}}}} \right) + \frac{1}{{1 - {x^2}}}} \right],
\nonumber\\
\label{Eq:fx}
\end{eqnarray}
%\end{widetext}
%%%%%%%%%%%%%%%%%%%%
leading to $f\left( x \right)\approx -3/2x$, for $x \gg 1$, which is the present case;
%%%%%%%%%%%%%%%%%%%%
\item The washout effect on $\varepsilon ^\alpha$ is controlled by $\eta\left(m^\alpha_{1eff}\right)$, which takes the form
%%%%%%%%%%%%%%%%%%%%
%\begin{widetext}
\begin{eqnarray}
&&
\eta \left( x \right) 
\nonumber\\
&&
= {\left( {\frac{{8.25 \times {{10}^{ - 3}}~{\rm{eV}}}}{x} + {{\left( {\frac{x}{{2 \times {{10}^{ - 4}}~{\rm{eV}}}}} \right)}^{1.16}}} \right)^{ - 1}},
\nonumber\\
\label{Eq:dilution}
\end{eqnarray}
%\end{widetext}
%%%%%%%%%%%%%%%%%%%%
where
%%%%%%%%%%%%%%%%%%%%
%\begin{widetext}
\begin{eqnarray}
m^\alpha_{1eff} = \frac{{{{\left( {m_D^\dag } \right)}_{1\alpha }}{{\left( {{m_D}} \right)}_{\alpha 1}}}}{{{M_1}}},
\label{Eq:EffectiveMass}
\end{eqnarray}
%\end{widetext}
%%%%%%%%%%%%%%%%%%%%
represents an effective mass;
\item For $10^9\lesssim M_1 \lesssim 10^{12}$ to be taken as our adopted range of $M_1$, the created lepton asymmetry $Y_L$, which becomes flavor-dependent, is calculated to be:
%%%%%%%%%%%%%%%%%%%%
%\begin{widetext}
\begin{eqnarray}
&&
Y_L\approx \frac{1}{g_\ast}\frac{12}{37}\left[ \left( {{\varepsilon ^e} + {\varepsilon ^\mu }} \right)\eta \left( {\frac{{417}}{{589}}\left( {{{\left| {{a_1}} \right|}^2} + {{\left| {{a_2}} \right|}^2}} \right)} \right)\right. 
\nonumber\\
&&
\quad
\left.+ {\varepsilon ^\tau }\eta \left( {\frac{{390}}{{589}}{{\left| {{a_3}} \right|}^2}} \right) \right],
\label{Eq:YL}
\end{eqnarray}
%\end{widetext}
%%%%%%%%%%%%%%%%%%%%
where $g_\ast$ is the effective thermodynamical number of the relativistic degree of freedom that is estimated to be 106.75 for the standard model at a cosmic temperature greater than 300 GeV and
%%%%%%%%%%%%%%%%%%%%
%\begin{widetext}
\begin{eqnarray}
{\varepsilon ^e} &=&  - \frac{{3{M_1}}}{{16\pi {v^2}}}\frac{{{\rm Im} \left[ {a_1^\ast {b_1}\left( {a_1^\ast {b_1} + a_2^\ast {b_2} + a_3^\ast {b_3}} \right)} \right]}}{{{{\left| {{a_1}} \right|}^2} + {{\left| {{a_2}} \right|}^2} + {{\left| {{a_3}} \right|}^2}}},
\nonumber\\
{\varepsilon ^\mu } &=&  - \frac{{3{M_1}}}{{16\pi {v^2}}}\frac{{{\rm Im} \left[ {a_2^\ast {b_2}\left( {a_1^\ast {b_1} + a_2^\ast {b_2} + a_3^\ast {b_3}} \right)} \right]}}{{{{\left| {{a_1}} \right|}^2} + {{\left| {{a_2}} \right|}^2} + {{\left| {{a_3}} \right|}^2}}},
\label{Eq:EachEpsilon}\\
{\varepsilon ^\tau } &=&  - \frac{{3{M_1}}}{{16\pi {v^2}}}\frac{{{\rm Im} \left[ {a_3^\ast {b_3}\left( {a_1^\ast {b_1} + a_2^\ast {b_2} + a_3^\ast {b_3}} \right)} \right]}}{{{{\left| {{a_1}} \right|}^2} + {{\left| {{a_2}} \right|}^2} + {{\left| {{a_3}} \right|}^2}}}.
\nonumber
\end{eqnarray}
%\end{widetext}
%%%%%%%%%%%%%%%%%%%%
\end{enumerate}
%%%%%%%%%%%%%%%%%%%%
The obtained $Y_L$ is related to the baryon asymmetry $Y_B$: $Y_B\approx -0.54Y_L$ and the final baryon-photon ratio $\eta_B(=(n_B-n_{\bar B})/n_\gamma)$ is estimated to be $\eta_B=7.04Y_B$.

To make theoretical predictions on $\eta_B$, let us choose the flavor-independent estimation of $\eta_B$, where $\eta_B$ is proportional to ${\rm Im}[( a_1^\ast{b_1} + a_2^\ast{b_2} + a_3^\ast{b_3})^2]$.  To see the dependence of $\eta_B$ on $\delta$ and $\phi$, we evaluate $\delta b_3$ appearing in Eq.(\ref{Eq:ApproximatedPhtonToBaryonBreaking}).  For the normal mass hierarchy, using the relations of
%%%%%%%%%%%%%%%%%%%%
%\begin{widetext}
\begin{eqnarray}
{a_2} &\approx& {\sigma _{\mu \mu }}\sqrt {{M_{\mu \mu }}},
\quad
{a_3} \approx {\sigma _{\tau \tau }}\sqrt {{M_{\tau \tau }}} 
\nonumber\\
{b_1} &=& {\sigma _{ee}}\sqrt {{M_{ee}}},
\quad
{b_2} = \frac{{{M_{e\mu }}}}{{{\sigma _{ee}}\sqrt {{M_{ee}}} }},
\nonumber\\
{b_3} &=& \frac{{{M_{e\tau }}}}{{{\sigma _{ee}}\sqrt {{M_{ee}}} }},
\label{Eq:SeeawMassMij}
\end{eqnarray}
%\end{widetext}
%%%%%%%%%%%%%%%%%%%%
where $\sigma_{ee, \mu\mu, \tau\tau}=\pm 1$ and $\sigma _{\tau \tau }=\sigma _{\mu \mu }$ is required for $\delta a_3$ to vanish at $\theta_{13}=0$, we find that Eq.(\ref{Eq:DeltaMetauMtautauByab}) yields
%%%%%%%%%%%%%%%%%%%%
%\begin{widetext}
\begin{eqnarray}
\delta {b_3} &\approx& \frac{{{\sigma _{ee}}{s_{13}}{c_{13}}}}{{{s_{12}}{c_{23}}}}\frac{{{e^{i\delta }}{{\tilde m}_3}}}{{\sqrt {{{\tilde m}_2}} }},
\label{Eq:da3dba}
\end{eqnarray}
%\end{widetext}
%%%%%%%%%%%%%%%%%%%%
from $\delta M_{e\tau}$ expressed in terms of $m_{2,3}$ \cite{DiracCPFormula}.  As a result,
%%%%%%%%%%%%%%%%%%%%
%\begin{widetext}
\begin{eqnarray}
a_1^\ast {b_1} + a_2^\ast {b_2} + a_3^\ast {b_3} \approx {\sigma _{ee}}{\sigma _{\mu \mu }}\frac{{{s_{13}}{c_{13}}}}{{{s_{12}}}}\sqrt {\frac{{{m_3}}}{{{m_2}}}} {m_3}{e^{i\left( {\delta  - \frac{\phi }{2}} \right)}},
\nonumber\\
&&
\label{Eq:BAU_m1m2}
\end{eqnarray}
%\end{widetext}
%%%%%%%%%%%%%%%%%%%%
is derived from Eq.(\ref{Eq:ApproximatedPhtonToBaryonBreaking}).  Since $\eta_B \propto{\rm Im}[( a_1^\ast{b_1} + a_2^\ast{b_2} + a_3^\ast{b_3})^2]$, we reach $\eta_B\propto\sin^2\theta_{13}\sin \left( {2\delta  - \phi } \right)$, which is the relation for the $a_1=0$ texture.  On the other hand, for the $b_1=0$ texture, we similarly find that $\eta_B\propto -\sin^2\theta_{13}\sin \left( {2\delta  - \phi } \right)$ from Eq.(\ref{Eq:ApproximatedPhtonToBaryonBreaking_b1=0}). Including $M_1$, we conclude that
%%%%%%%%%%%%%%%%%%%%
%\begin{widetext}
\begin{eqnarray}
\eta_B\propto \xi_B M_1 \sin^2\theta_{13}\sin \left( {2\delta  - \phi } \right),
\label{Eq:eta_B_delta_phi}
\end{eqnarray}
%\end{widetext}
%%%%%%%%%%%%%%%%%%%%
serves as a good prediction of $\eta_B$, where $\xi_B =1$ ($\xi_B =-1$) for the $a_1=0$ ($b_1=0$) texture.

We also derive the relation between $\delta$ and $\phi$ from the $i=\mu$ part of Eq.(\ref{Eq:ScalingRule}) equivalent to Eq.(\ref{Eq:Sin2SC}), which can be rephrased in terms of $m_{2,3}$ as follows:
%%%%%%%%%%%%%%%%%%%%
%\begin{widetext}
\begin{eqnarray}
&&
\left( {A{s_{23}}{c_{13}}{c_{23}}{c_{13}} + B{t_{23}}s_{23}^2c_{13}^2} \right){m_3}{e^{ - i\phi }}
\nonumber\\
&&
\quad
= \left[ {A\left( {{c_{23}}{c_{12}} - {s_{23}}{s_{12}}\tilde s_{13}^\ast } \right)\left( {{s_{23}}{c_{12}} + {c_{23}}{s_{12}}\tilde s_{13}^\ast } \right)} \right.
\nonumber\\
&&
\quad
\left. { - B{t_{23}}{{\left( {{c_{23}}{c_{12}} - {s_{23}}{s_{12}}\tilde s_{13}^\ast } \right)}^2}} \right]{m_2}.
\label{Eq:MmumuMmutauTom2m3}
\end{eqnarray}
%\end{widetext}
%%%%%%%%%%%%%%%%%%%%
Therefore, we find that $\phi=0$ at $\theta_{13}=0$. For $\theta_{13}\neq 0$, the right-handed side of Eq.(\ref{Eq:MmumuMmutauTom2m3}) can be approximated to be $( {1 - {t_{23}}{t_{12}}\tilde s_{13}^\ast })t_{23}^3c_{12}^2{m_2}$, from which we derive
%%%%%%%%%%%%%%%%%%%%
%\begin{widetext}
\begin{equation}
\tan \phi  \approx  - {t_{23}}{t_{12}}{s_{13}}\sin \delta,
\label{Eq:delta-phi-2}
\end{equation}
%\end{widetext}
%%%%%%%%%%%%%%%%%%%%
numerically leading to $\tan\phi\approx -0.1 \sin\delta$ for the observed data.  We expect that the magnitude of $\phi$ is at most 0.1.

Since we would like to discuss effects of the CP-violating Dirac phase $\delta$ on the creation of $Y_L$, we may consider the renormalization effects that modify the magnitude of $\delta$ when $\delta$ is promoted into $Y_L$. It has been discussed that the renormalization effect is rather insignificant for neutrinos in the normal mass hierarchy \cite{RGE}, where we reside now.  
%%%%%%%%%%%%%%%%%%%% FIG %%%%%%%%%%%%%%%%%%%%
%%%%%%%%%%%%%%%%%%%%
%%%%%%%%%%%%%%%%%%%%
%--------------------------------------------------------------------
\begin{figure}[t]
\begin{center}
%\includegraphics*[20mm,80mm][200mm,265mm]{phase.eps}
%\includegraphics*[20mm,90mm][200mm,265mm]{phase.eps}
%80==>90 title moves closer to figure
%\includegraphics*[20mm,90mm][200mm,265mm]{phase.eps}
%\includegraphics*[20mm,90mm][200mm,235mm]{phase.eps}
%265==>235 figure moves upward
\includegraphics*[13mm,235mm][150mm,291mm]{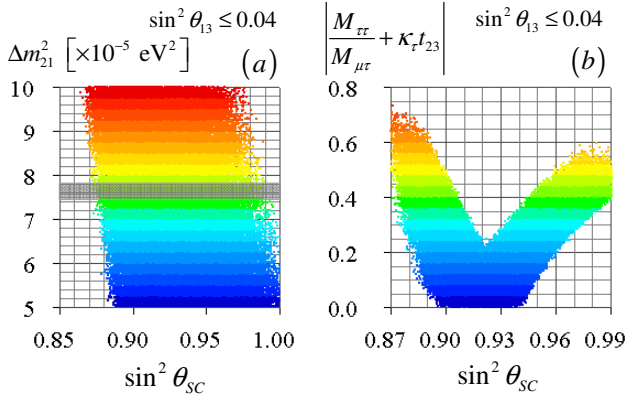}
\caption{(a) $\Delta m_{21}^2$ as a function of $\sin^2\theta_{SC}$, where the grey rectangle denotes the experimentally allowed region of $\Delta m_{21}^2$, and (b) $\left| \left(M_{\tau \tau }/M_{\mu \tau }\right) + \kappa _\tau t_{23}\right|$ as a function of $\sin^2\theta_{SC}$, where $\sin^2\theta_{SC}$ is restricted to reproduce the observed $\Delta m_{21}^2$, neither of which depends on the texture type.}
\label{Fig:vsS2AtmDMtautau}
\end{center}
\end{figure}
%--------------------------------------------------------------------
\begin{figure}[t]
\begin{center}
%\includegraphics*[20mm,80mm][200mm,265mm]{phase.eps}
%\includegraphics*[20mm,90mm][200mm,265mm]{phase.eps}
%80==>90 title moves closer to figure
%\includegraphics*[20mm,90mm][200mm,265mm]{phase.eps}
%\includegraphics*[20mm,90mm][200mm,235mm]{phase.eps}
%265==>235 figure moves upward
\includegraphics*[13mm,228mm][150mm,291mm]{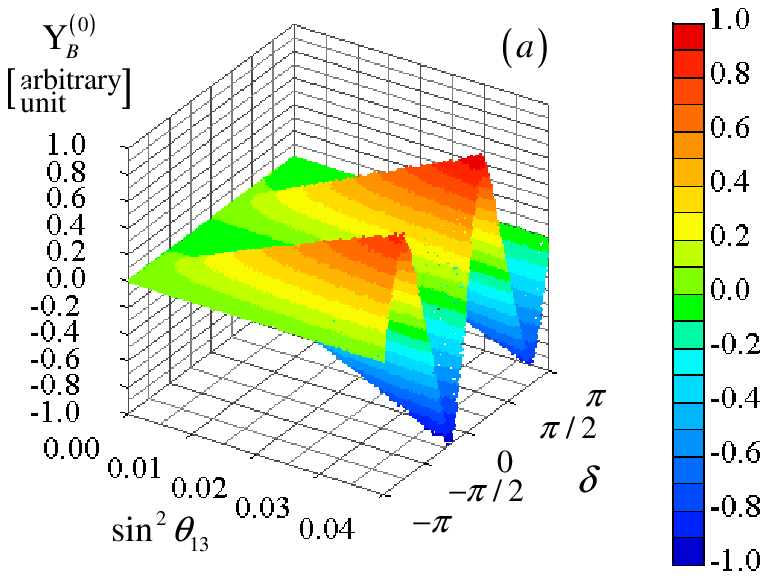}
\includegraphics*[13mm,227mm][150mm,296mm]{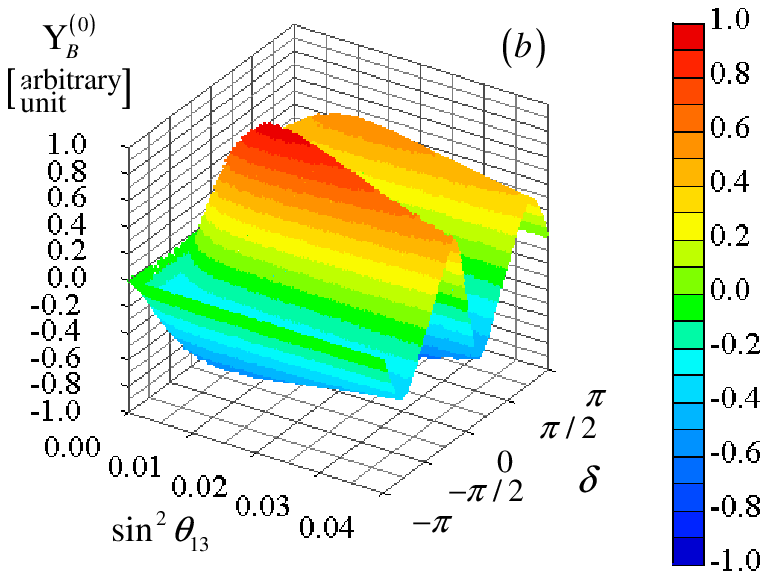}
\caption{$Y^{(0)}_B$, the appropriately normalized $Y_B/M_1$, as a function of $\sin^2\theta_{13}$ and $\delta$ for (a) the $a_1=0$ texture and (b) the $b_1=0$ texture.}
\label{Fig:vsSin13_2DeltaNegativeBAU}
\end{center}
\end{figure}
%--------------------------------------------------------------------
\begin{figure}[t]
\begin{center}
%\includegraphics*[20mm,80mm][200mm,265mm]{phase.eps}
%\includegraphics*[20mm,90mm][200mm,265mm]{phase.eps}
%80==>90 title moves closer to figure
%\includegraphics*[20mm,90mm][200mm,265mm]{phase.eps}
%\includegraphics*[20mm,90mm][200mm,235mm]{phase.eps}
%265==>235 figure moves upward
\includegraphics*[8mm,235mm][150mm,290mm]{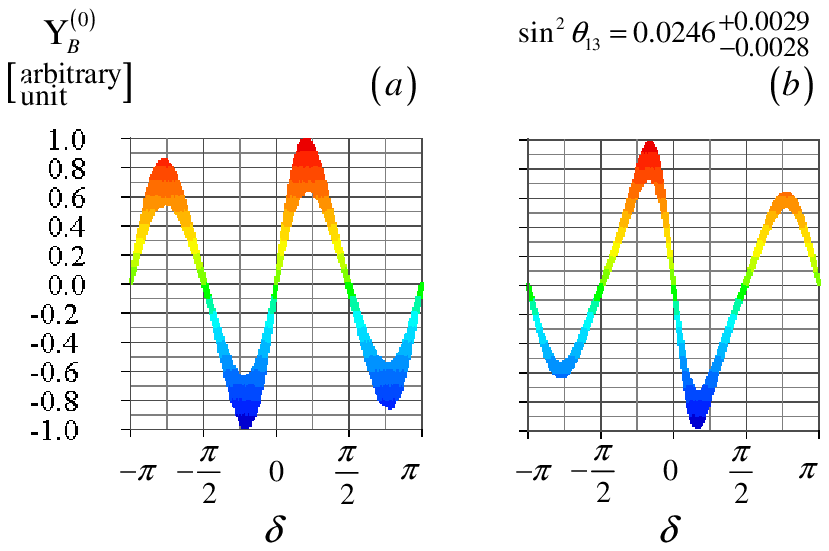}
\caption{$Y^{(0)}_B$, the appropriately normalized $Y_B/M_1$, as a function of $\delta$ for (a) the $a_1=0$ texture and (b)  the $b_1=0$ texture.}
\label{Fig:vsDeltaNegativeBAUBareFixedSin13}
\includegraphics*[8mm,235mm][150mm,304mm]{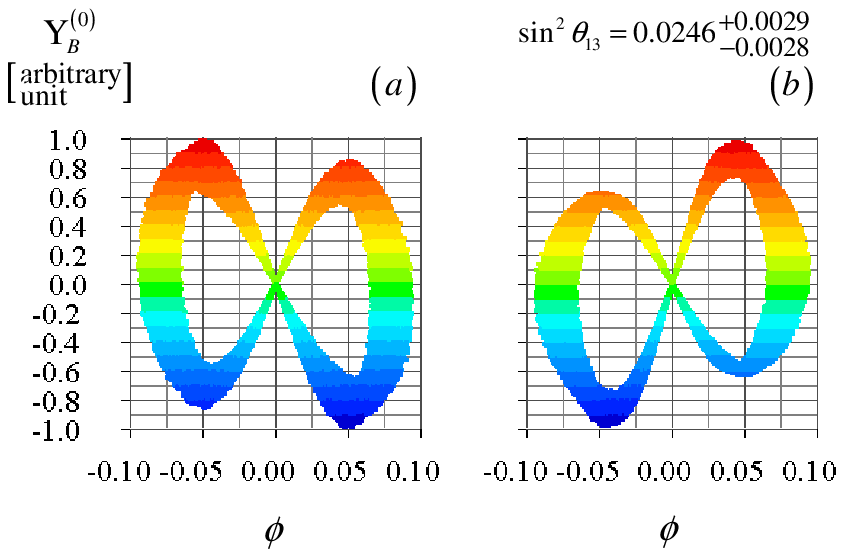}
\caption{The same as in FIG.{\ref{Fig:vsDeltaNegativeBAUBareFixedSin13}} but for $\phi$.}
\label{Fig:vsPhiNegativeBAUBareFixedSin13}
\end{center}
\end{figure}
%--------------------------------------------------------------------
\begin{figure}[t]
\begin{center}
%\includegraphics*[20mm,80mm][200mm,265mm]{phase.eps}
%\includegraphics*[20mm,90mm][200mm,265mm]{phase.eps}
%80==>90 title moves closer to figure
%\includegraphics*[20mm,90mm][200mm,265mm]{phase.eps}
%\includegraphics*[20mm,90mm][200mm,235mm]{phase.eps}
%265==>235 figure moves upward
\includegraphics*[8mm,235mm][150mm,290mm]{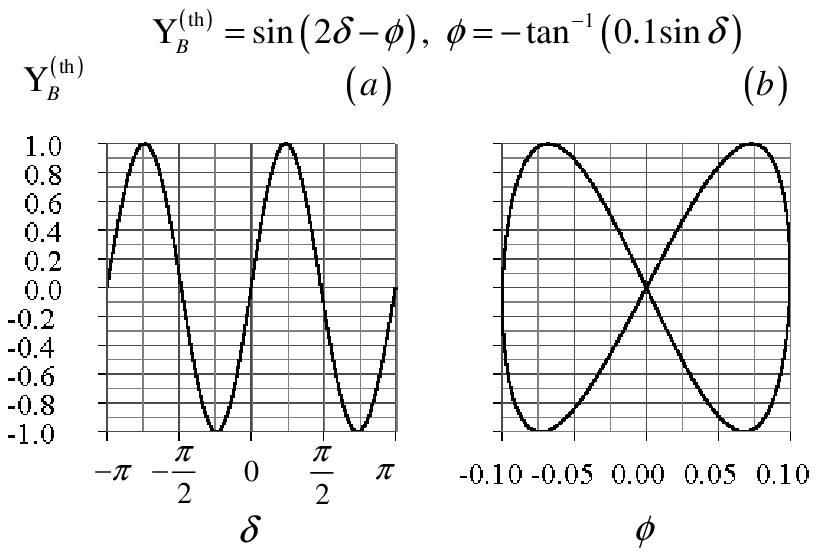}
\caption{$Y^{({\rm th})}_B=\sin(2\delta-\phi)$ with $\phi=-\tan^{-1}(0.1\sin\delta)$ as a function of (a) $\delta$ and (b) $\phi$ for the $a_1=0$ texture, where figures for the $b_1=0$ texture are obtained by plotting $Y^{({\rm th})}_B=-\sin(2\delta-\phi)$ with $\phi$ = $-\tan^{-1}(0.1\sin\delta)$.}
\label{Fig:vsDeltaPhiBAUbareSinCurve}
\includegraphics*[8mm,235mm][150mm,296mm]{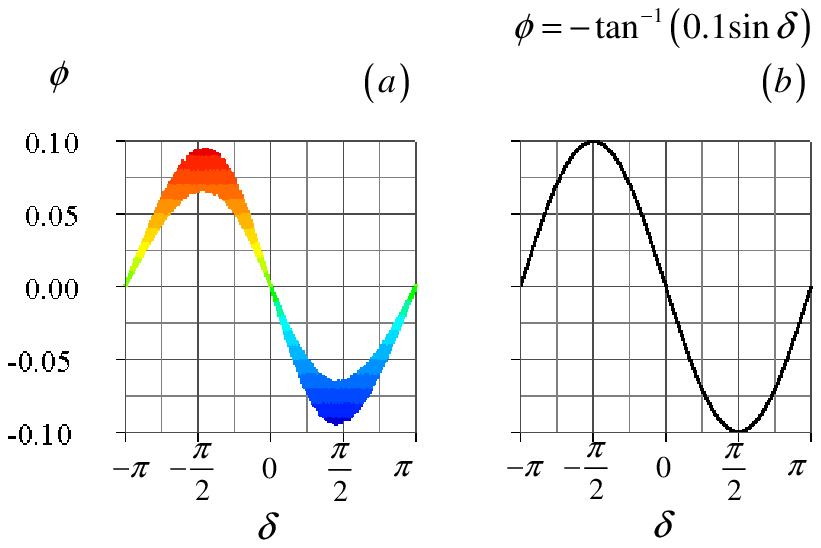}
\caption{$\phi$ as a function of $\delta$ for (a) numerical calculation of $\phi$ and (b) $\phi=-\tan^{-1}\left(0.1\sin\delta\right)$}
\label{Fig:vsDeltaPhiFixedSin13}
\end{center}
\end{figure}
%--------------------------------------------------------------------
\begin{figure}[t]
\begin{center}
%\includegraphics*[20mm,80mm][200mm,265mm]{phase.eps}
%\includegraphics*[20mm,90mm][200mm,265mm]{phase.eps}
%80==>90 title moves closer to figure
%\includegraphics*[20mm,90mm][200mm,265mm]{phase.eps}
%\includegraphics*[20mm,90mm][200mm,235mm]{phase.eps}
%265==>235 figure moves upward
\includegraphics*[13mm,222mm][150mm,289mm]{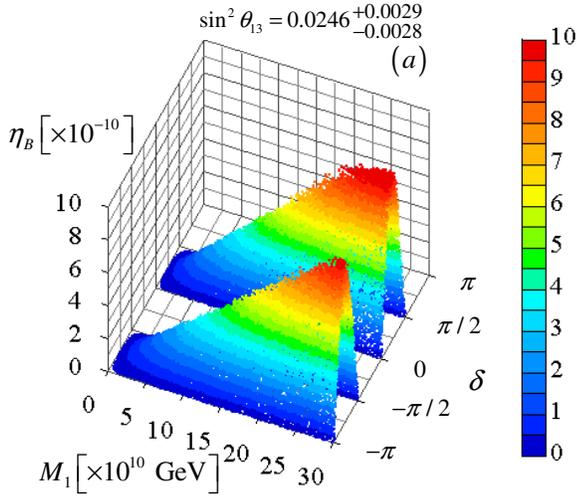}
\includegraphics*[13mm,222mm][150mm,292mm]{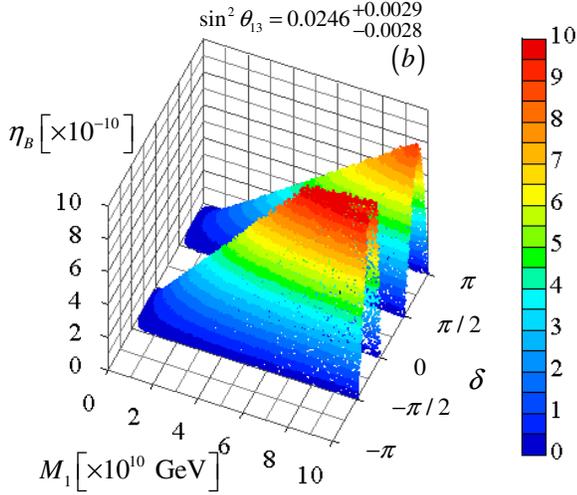}
\caption{$\eta_B$ as a function of $M_1$ and $\delta$ for (a) the $a_1=0$ texture and (b) the $b_1=0$ texture.}
\label{Fig:vsM1DeltaBAUFixedSin13}
\end{center}
\end{figure}
%--------------------------------------------------------------------
\begin{figure}[t]
\begin{center}
%\includegraphics*[20mm,80mm][200mm,265mm]{phase.eps}
%\includegraphics*[20mm,90mm][200mm,265mm]{phase.eps}
%80==>90 title moves closer to figure
%\includegraphics*[20mm,90mm][200mm,265mm]{phase.eps}
%\includegraphics*[20mm,90mm][200mm,235mm]{phase.eps}
%265==>235 figure moves upward
\includegraphics*[8mm,233mm][150mm,290mm]{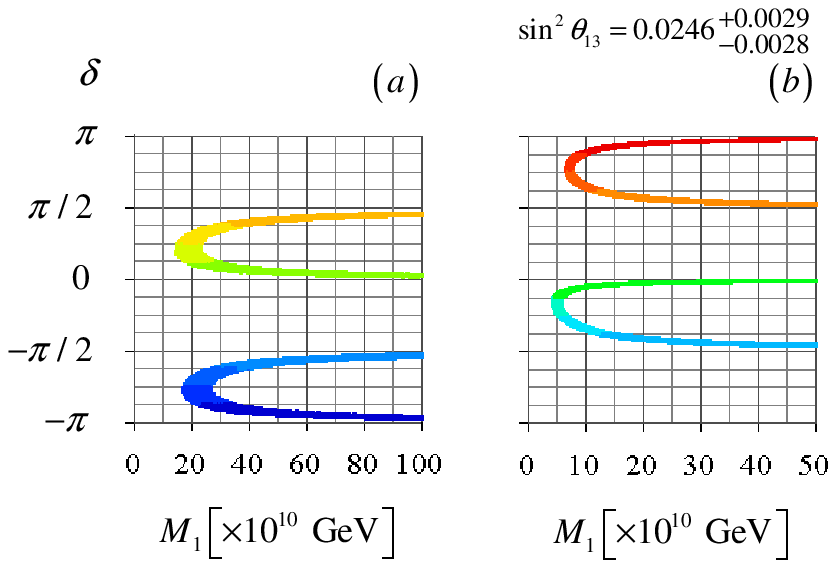}
\caption{$\delta$ as a function of $M_1$ to reproduce the observed $\eta_B$ for (a) the $a_1=0$ texture and (b) the $b_1=0$ texture.}
\label{Fig:vsM1Delta}
\includegraphics*[8mm,233mm][150mm,295mm]{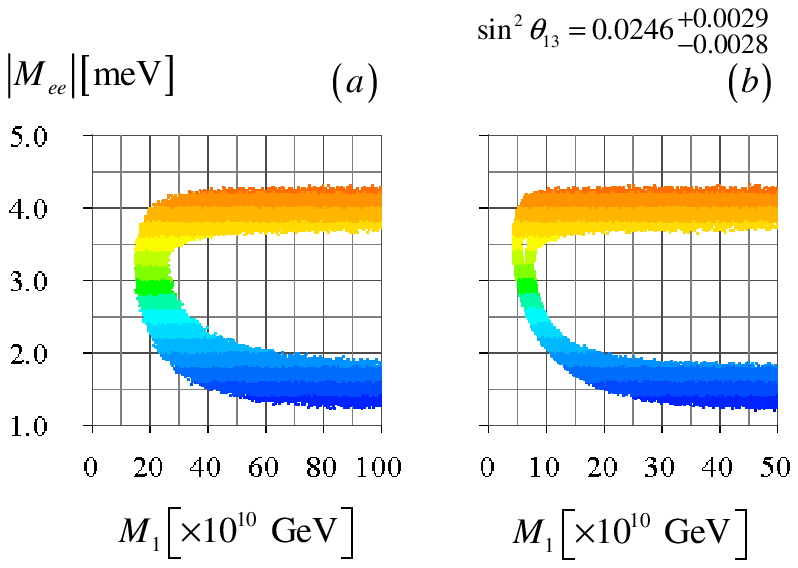}
\caption{The same as in FIG.{\ref{Fig:vsM1Delta}} but for $\left|M_{ee}\right|$.}
\label{Fig:vsM1Mee}
\end{center}
\end{figure}
%%%%%%%%%%%%%%%%%%%%
%%%%%%%%%%%%%%%%%%%%

\section{\label{sec:numerical} Numerical Analysis}
We perform numerical calculation of $\eta_B$ by adopting the following parameters obtained from neutrino oscillations \cite{NuData}: 
%%%%%%%%%%%%%%%%%%%%
%\begin{widetext}
\begin{eqnarray}
&&
\Delta m^2_{21} ~[10^{-5}~{\rm eV}^2]  = 7.62\pm 0.19,
\nonumber\\
&&
\Delta m^2_{31} ~[10^{-3}~{\rm eV}^2] = 2.55
{\footnotesize
{\begin{array}{*{20}c}
   { + 0.06}  \\
   { - 0.09}  \\
\end{array}}
},
\label{Eq:NuDataMass}\\
&&
\sin ^2 \theta _{12}  = 0.320
{\footnotesize
{\begin{array}{*{20}c}
   { + 0.016}  \\
   { - 0.017}  \\
\end{array}}
},
\quad
\sin ^2 \theta _{23} =0.427
{\footnotesize
{\begin{array}{*{20}c}
   { + 0.034}  \\
   { - 0.027}  \\
\end{array}}
},
\nonumber\\
&&
\sin ^2 \theta _{13} =0.0246
{\footnotesize
 {\begin{array}{*{20}c}
   { + 0.0029}  \\
   { - 0.0028}  \\
\end{array}}
},
\label{Eq:NuDataAngle}
\end{eqnarray}
%\end{widetext}
%%%%%%%%%%%%%%%%%%%%
where $\Delta m^2_{ij}=m^2_i-m^2_j$ for $m_i$ specifying a mass of $\nu_i$ ($i=1,2,3$).  There is another similar analysis that has reported the slightly smaller values of $\sin^2\theta_{23} = 0.365-0.410$ \cite{NuData2}.  The created $\eta_B$ should be consistent with the WMAP observed data \cite{WMAP} of
%%%%%%%%%%%%%%%%%%%%
%\begin{widetext}
\begin{eqnarray}
\eta_B = \left(6.2\pm 0.15\right)\times 10^{-10}.
\label{Eq:WMAP}
\end{eqnarray}
%\end{widetext}
%%%%%%%%%%%%%%%%%%%%
 To study the dependence of $\eta_B$ on $\theta_{13}$, $\delta$ and $\phi$, we use $Y^{(0)}_B$ as an appropriately normalized $Y_B/M_1$, which becomes 
%%%%%%%%%%%%%%%%%%%%
%\begin{widetext}
\begin{eqnarray}
Y^{(0)}_B\propto \xi_B \sin^2\theta_{13}\sin \left( {2\delta  - \phi } \right).
\label{Eq:Y0_B}
\end{eqnarray}
%\end{widetext}
%%%%%%%%%%%%%%%%%%%%

We have searched acceptable parameter regions by changing $M_1$ up to $10^{12}$ GeV for the $a_1=0$ texture and up to $5\times 10^{11}$ GeV for the $b_1=0$ texture to see how BAU is created and how BAU depends on $\theta_{13}$ and $\delta$.  The neutrino masses, mixing angles and $\eta_B$ are constrained by their experimental data, Eqs.(\ref{Eq:NuDataMass}), (\ref{Eq:NuDataAngle}) and (\ref{Eq:WMAP}), unless they are specified.  Our theoretical predictions, which have been obtained by using certain approximations, are to be compared with numerical results obtained without such approximations.  The results of our numerical analysis are shown in FIG.\ref{Fig:vsS2AtmDMtautau}-FIG.\ref{Fig:vsM1Mee}: 
%%%%%%%%%%%%%%%%%%%%
\begin{enumerate}
%%%%%%%%%%%%%%%%%%%%
\item In FIG.\ref{Fig:vsS2AtmDMtautau} (a) for $\sin^2\theta_{13}\leq 0.04$, the angle $\theta_{SC}$ should satisfy $0.87\lesssim\sin^2\theta_{SC}\lesssim 0.99$ to cover the observed range of $\Delta m^2_{21}$.
\item FIG.\ref{Fig:vsS2AtmDMtautau} (b) for $\sin^2\theta_{13}\leq 0.04$ shows what extent the general scaling rule of ${M_{\tau \tau }}/{M_{\mu \tau }} =  - {\kappa _\tau }{t_{23}}$ is satisfied and this scaling rule turns out to be satisfied within 70\%.
\item FIG.\ref{Fig:vsSin13_2DeltaNegativeBAU} shows how $Y^{(0)}_B$, the appropriately normalized $Y_B/M_1$, evolves with $\sin^2\theta_{13}$ and $\delta$: 
%%%%%%%%%%%%%%%%%%%%
\begin{enumerate}
%%%%%%%%%%%%%%%%%%%%
\item The feature that $Y^{(0)}_B$ increases as $\sin^2\theta_{13}$ increases appears for both textures although it starts decreasing around $\sin^2\theta_{13}\approx 0.015$ for the $b_1=0$ texture and the predicted proportionality of $Y^{(0)}_B$ to $\sin^2\theta_{13}$ in Eq.(\ref{Eq:Y0_B}) is more visible for the $a_1=0$ texture;
\item The oscillating $Y^{(0)}_B$ with $\delta$ is compatible with the prediction of Eq.(\ref{Eq:Y0_B}), which is also compared with the results of FIG.\ref{Fig:vsDeltaNegativeBAUBareFixedSin13} and FIG.\ref{Fig:vsPhiNegativeBAUBareFixedSin13};
\item Leptogenesis starts creating BAU if $0\lesssim\delta\lesssim \pi/2$ (mod $\pi$) for the $a_1=0$ texture and if $-\pi/2\lesssim\delta\lesssim 0$ (mod $\pi$) for the $b_1=0$ texture.
%%%%%%%%%%%%%%%%%%%%
\end{enumerate}
%%%%%%%%%%%%%%%%%%%%
\item When $\theta_{13}$ is restricted to the observed values, FIG.\ref{Fig:vsDeltaNegativeBAUBareFixedSin13} and FIG.\ref{Fig:vsPhiNegativeBAUBareFixedSin13} can be used to examine the oscillating behavior found in FIG.\ref{Fig:vsSin13_2DeltaNegativeBAU}. The gross features of the oscillating behavior of $Y^{(0)}_B$ in FIG.\ref{Fig:vsDeltaNegativeBAUBareFixedSin13} and of the Lissajous-like behavior of $Y^{(0)}_B$ in FIG.\ref{Fig:vsPhiNegativeBAUBareFixedSin13} can be accounted by the prediction of Eq.(\ref{Eq:Y0_B}) as long as $\tan\phi\approx -0.1 \sin\delta$ from Eq.(\ref{Eq:delta-phi-2}) is used to calculate $\phi$.  Namely, two graphs of $Y^{({\rm th})}_B=\sin(2\delta-\phi)$ with $\tan\phi=-0.1\sin\delta$ plotted in FIG.\ref{Fig:vsDeltaPhiBAUbareSinCurve} for the $a_1=0$ texture depict similar shapes to those in FIG.\ref{Fig:vsDeltaNegativeBAUBareFixedSin13} (a) and FIG.\ref{Fig:vsPhiNegativeBAUBareFixedSin13} (a). For the $b_1=0$ texture, $Y^{({\rm th})}_B=-\sin(2\delta-\phi)$ with $\tan\phi$ = $-0.1\sin\delta$ similarly accounts for the behavior of $Y^{(0)}_B$.
\item FIG.\ref{Fig:vsDeltaPhiFixedSin13} compares the result of the calculated $\phi$ as a function of $\delta$ with  our prediction of $\phi=-\tan^{-1}\left(0.1\sin\delta\right)$, where we understand that our prediction plotted in FIG.\ref{Fig:vsDeltaPhiFixedSin13} (b) can simulate $\phi$. 
\item The minimum value of $M_1$ to reproduce the observed $\eta_B$ can be determined by FIG.\ref{Fig:vsM1DeltaBAUFixedSin13} and, more explicitly, by FIG.\ref{Fig:vsM1Delta} to be:
%%%%%%%%%%%%%%%%%%%%
\begin{enumerate}
%%%%%%%%%%%%%%%%%%%%
\item $1.5\times 10^{11}$ GeV ($1.8\times 10^{11}$ GeV) if $0 < \delta < \pi/2$ ($-\pi < \delta < -\pi/2$) for the $a_1=0$ texture;
\item $6.5\times 10^{10}$ GeV ($4.5\times 10^{10}$ GeV) if $\pi/2 < \delta < \pi$ ($-\pi/2 < \delta < 0$) for the $b_1=0$ texture.
%%%%%%%%%%%%%%%%%%%%
\end{enumerate}
%%%%%%%%%%%%%%%%%%%%
Since BAU inversely depends on $\left|a^2_1\right|+\left|a^2_2\right|+\left|a^2_3\right|$, larger amount of $\eta_B$ is expected for $\left|a^2_{1,2,3}\right|\ll \left|b^2_{2,3}\right|$ corresponding to the $b_1=0$ texture that certainly allows $M_1$ to take smaller values as in (b). 
\item When $\eta_B$ and $\theta_{13}$ are consistent with the observed values, the correlation of $\delta$ with $M_1$ is shown in FIG.\ref{Fig:vsM1Delta}, where its behavior can also be explained by the prediction of $\eta_B\propto \xi_B M_1 \sin^2\theta_{13}\sin (2\delta-\phi)$ with $\tan\phi\approx -0.1\sin\delta$.
%%%%%%%%%%%%%%%%%%%%
\begin{enumerate}
%%%%%%%%%%%%%%%%%%%%
\item For the $a_1=0$ texture ($\xi_B =1$), $\delta$ tends to approach 0 and $\pi/2$ as $M_1$ increases and this behavior is consistent with the prediction of the factor $M_1 \sin (2\delta-\phi)$ with $\left|\phi\right|\lesssim 0.1$, which tends to stay at the appropriate value corresponding to the observed $\eta_B$ and which requires that $\left|\sin (2\delta-\phi)\right|$ gets smaller as $M_1$ gets larger and either $\delta\approx 0$ or $\delta\approx \pi/2$ (mod $\pi$) is a target value.
\item For the $b_1=0$ texture ($\xi_B =-1$), the same reasoning leads to the behavior that $\delta$ tends to approach $\pi/2$ and $\pi$ (mod $\pi$) as $M_1$ increases.
\item Near the threshold to start creating the observed $\eta_B$, $\delta$ points to the value such that $\left|\sin (2\delta-\phi)\right|$ with $\left|\phi\right|\lesssim 0.1$ is nearly maximal and $\delta\approx \pi/4$ (mod $\pi$) is selected for the $a_1=0$ texture as can be read off from FIG.\ref{Fig:vsM1Delta} (a) while $\delta\approx -\pi/4$ (mod $\pi$) is selected for the $b_1=0$ texture as in FIG.\ref{Fig:vsM1Delta} (b).
%%%%%%%%%%%%%%%%%%%%
\end{enumerate}
%%%%%%%%%%%%%%%%%%%%
\item $\left|M_{ee}\right|$ to be measured by neutrinoless double beta decay \cite{DoubleBeta} is computed to show FIG.\ref{Fig:vsM1Mee} when $\eta_B$ and $\theta_{13}$ are consistent with the observed values, which describes the correlation of $\left|M_{ee}\right|$ with $M_1$.  We find that
%%%%%%%%%%%%%%%%%%%%
\begin{enumerate}
%%%%%%%%%%%%%%%%%%%%
\item 1.2 meV $\lesssim\left|M_{ee}\right|\lesssim$ 4.0 meV;
\item $\left|M_{ee}\right|$ starting around 3 meV increases up to around 4 meV or decreases down to around 1.5 meV as $M_1$ increases.
%%%%%%%%%%%%%%%%%%%%
\end{enumerate}
%%%%%%%%%%%%%%%%%%%%
The behavior of $\left| M_{ee} \right|$ is consistent with the known estimation of $\left|M_{ee}\right| = \left| c_{13}^2s_{12}^2{m_2}e^{i\left( {\phi  - 2\delta } \right)} + s_{13}^2{m_3} \right|$ once the constraint that $M_1\sin 2\delta$ is nearly constant is taken into account. For the $a_1=0$ texture, at $\delta=0$, $\delta=\pi/4$ and  $\delta=\pi/2$ selected as the key values in list \#6, scales of 4.3 meV, 3.3 meV (around the threshold) and 1.7 meV can be, respectively, calculated for $\phi=0$.  Then, we expect that $\left|M_{ee}\right|$ starting around 3.3 meV increases toward 4.3 meV or decreases toward 1.7 meV as shown in FIG.\ref{Fig:vsM1Mee} (a).  The same explanation is possible for the $b_1=0$ texture in FIG.\ref{Fig:vsM1Mee} (b).
%%%%%%%%%%%%%%%%%%%%
\end{enumerate}
%%%%%%%%%%%%%%%%%%%%

\section{\label{sec:summary} Summary}
We have found minimal seesaw models compatible with the generalized scaling ansatz of Eq.(\ref{Eq:ScalingRule}).  The angle $\theta_{SC}$ is determined to be ${\sin ^2}{\theta _{SC}} = c_{23}^2\left( {{M_{\mu \tau }} + {t_{23}}{M_{\mu \mu }}} \right)/\left[ {\left( {1 - t_{23}^2} \right){M_{\mu \tau }} + {t_{23}}{M_{\mu \mu }}} \right]$, which satisfies $0.87 \lesssim \sin^2\theta_{SC}\lesssim 0.99$, where $\Delta m^2_{21}$ can stay in the observed range.  The first seesaw texture is described by
%%%%%%%%%%%%%%%%%%%%
%\begin{widetext}
\begin{eqnarray}
{m_D} = \left( {\begin{array}{*{20}{c}}
{\sqrt {{M_1}} {a_1}}&{\sqrt {{M_2}} {b_1}}\\
{\sqrt {{M_1}} {a_2}}&{\sqrt {{M_2}} {b_2}}\\
{\sqrt {{M_1}} \left( { - {t_{23}}{a_2}+\delta a_3} \right)}&{\sqrt {{M_2}} \left( { - {t_{23}}{b_2}}+\delta b_3 \right)}
\end{array}} \right),
\nonumber\\
\label{Eq:Summary_Solution_mD_1}
\end{eqnarray}
%\end{widetext}
%%%%%%%%%%%%%%%%%%%%
where $\sin^2\theta_{SC}=0$ is derived.  The second seesaw textures consist of
%%%%%%%%%%%%%%%%%%%%
%\begin{widetext}
\begin{eqnarray}
{m_D} = \left( {\begin{array}{*{20}{c}}
{0}&{\sqrt {{M_2}} {b_1}}\\
{\sqrt {M_1}} {a_2}&{\sqrt {{M_2}} {b_2}}\\
{\sqrt {M_1}} \left(a_2/t_{23}+\delta a_3\right)&{\sqrt {{M_2}} \left( { - {t_{23}}{b_2}}+\delta b_3 \right)}
\end{array}} \right),
\nonumber\\
\label{Eq:Summary_Solution_mD_2}
\end{eqnarray}
%\end{widetext}
%%%%%%%%%%%%%%%%%%%%
where ${\sin ^2}{\theta _{SC}} = {\left(a_2/t_{23}\right)^2}/\left[{\left(a_2/t_{23}\right)^2 + \left(t_{23}b_2\right)^2}\right]$, and
%%%%%%%%%%%%%%%%%%%%
%\begin{widetext}
\begin{eqnarray}
{m_D} = \left( {\begin{array}{*{20}{c}}
{\sqrt {{M_1}} {a_1}}&{0}\\
{\sqrt {M_1}} {a_2}&{\sqrt {{M_2}} {b_2}}\\
{\sqrt {M_1} \left( { - {t_{23}}{a_2}}+\delta a_3 \right)}&{\sqrt {M_2}} \left(b_2/t_{23}+\delta b_3\right)
\end{array}} \right),
\nonumber\\
\label{Eq:Summary_Solution_mD_3}
\end{eqnarray}
%\end{widetext}
%%%%%%%%%%%%%%%%%%%%
where ${\sin ^2}{\theta _{SC}} = {\left(b_2/t_{23}\right)^2}/\left[{\left(t_{23}a_2\right)^2+\left(b_2/t_{23}\right)^2}\right]$.  However, the second textures in the inverted mass hierarchy with $m_3=0$ cannot be connected to those at $\theta_{13}=0$ but are connected to the first texture with either $a_1=0$ or $b_1=0$. Therefore, the second textures should yield the normal mass hierarchy.

It is demonstrated that BAU vanishes for the second textures in the exact scaling limit. For the $a_1=0$ texture, the onset of CP-violations and $\theta_{13} \neq 0$ is signaled by the nonvanishing $\delta b_3$. Namely, BAU depends on $a_2^\ast \delta {b_3}$, the CP-violating Dirac phase is approximated to be $\arg \left( a_2^{\ast 2}b_1\delta b_3 \right)$ and $\theta_{13}$ is evaluated to give $\tan 2{\theta _{13}} \approx 2c_{23}s_{23}^2\left| {{{{b_1}\delta {b_3}}}/{{a_2^2}}} \right|$. A similar conclusion is derived for the $b_1=0$ texture.

Our main prediction is $\eta_B\propto M_1\sin^2\theta_{13}$$\sin ( 2\delta$$ - \phi)$ for the $a_1=0$ texture and $\eta_B$ $\propto$ $-M_1 \sin^2\theta_{13}\sin \left( {2\delta  - \phi } \right)$ for the $b_1=0$ texture together with $\tan\phi\approx - {t_{23}}{t_{12}}{s_{13}}\sin\delta$.  The  features of $\eta_B$ found in the numerical analysis based on the flavor-dependent leptogenesis turn out to be consistent with our predictions based on the simplified flavor-independent one: 
%%%%%%%%%%%%%%%%%%%%
\begin{enumerate}
%%%%%%%%%%%%%%%%%%%%
\item The proportionality of $\eta_B$ to $\sin^2\theta_{13}$ shows up and is more visible for the $a_1=0$ texture;
\item When $\theta_{13}$ is constrained to be the observed value, the oscillating $\eta_B$ with $\delta$ and $\phi$ is well observed for both textures and is consistent with the prediction once the relation of $\tan\phi\approx -0.1\sin\delta$ is included. 
%%%%%%%%%%%%%%%%%%%%
\end{enumerate}
%%%%%%%%%%%%%%%%%%%%

Leptogenesis starts creating the sufficient amount of BAU compatible with the WMAP observation if 
%%%%%%%%%%%%%%%%%%%%
\begin{enumerate}
%%%%%%%%%%%%%%%%%%%%
\item $M_1\gtrsim 1.5\times 10^{11}$ GeV with $0 < \delta < \pi/2$ and $M_1\gtrsim 1.8\times 10^{11}$ GeV with $-\pi < \delta < -\pi/2$ for the $a_1=0$ texture;
\item $M_1\gtrsim 6.5\times 10^{10}$ GeV with $\pi/2 < \delta < \pi$ and $M_1\gtrsim 4.5\times 10^{10}$ GeV with $-\pi/2 < \delta < 0$ for the $b_1=0$ texture.
%%%%%%%%%%%%%%%%%%%%
\end{enumerate}
%%%%%%%%%%%%%%%%%%%%
The $M_1$ dependence of $\delta$ to reproduce the observed $\eta_B$ is determined by the constraint that $M_1\sin(2\delta-\phi)$ is nearly constant.  For $M_1$ to initiate leptogenesis, $\left|\sin (2\delta-\phi)\right|\approx 1$ with $\left|\phi\right|\lesssim 0.1$ is required and $\delta$ is predicted to be near $\pi/4$ (mod $\pi$).  On the other hand, for the larger $M_1$, $\sin (2\delta-\phi)\approx 0$ is required to lead to $\delta\approx 0, \pi/2$ (mod $\pi$).  These two values are smoothly connected from $\delta\approx \pi/4$ (mod $\pi$) in the intermediate range of $M_1$. For $\left|M_{ee}\right|$, it is found that 1.2 meV $\lesssim\left|M_{ee}\right|\lesssim$ 4.0 meV.  The $M_1$ dependence of $\left|M_{ee}\right|$, which is a function of $2\delta-\phi$, is understood in a similar way.

\vspace{3mm}
%%%%%%%%%%%%%%%%%%%%%%%%%%%%%%%%%%%%%%%%%%%%%%%%%%%%%%%%%%%%%%%%%%%%%%%%%%%%%%%
\noindent
\centerline{\small \bf ACKNOWLEGMENTS}

The author would like to thank T. Kitabayashi for reading manuscript and useful comments.
%%%%%%%%%%%%%%%%%%%%%%%%%%%%%%%%%%%%%%%%%%%%%%%%%%%%%%%%%%%%%%%%%%%%%%%%%%%%%%%
%%-------------------------------------------------
%% References
%%-------------------------------------------------
%%%%%%%%%%%%%%%%%%%%%%%%%%%%%%%%%%%%%%%%%%%%%%

%%%%%%%%%%%%%%%%%%%%%%%%%%%%%%%%%%%%%%%%%%%%%%%%%%%%%%%%%%%%%%%%%%%%%%%%%%%%%%%%
\end{document}